# L'accessibilité des E-services aux personnes non-voyantes : difficultés d'usage et recommandations

*Françoise SANDOZ-GUERMOND*

ICTT - INSA
21, avenue Jean Capelle
69621 Villeurbanne Cedex France
françoise.sandoz-guermond@insa-lyon.fr

*Marc-Eric BOBILLER-CHAUMON*

ICTT - ECL
36, avenue Guy de Collongue
BP 163 – F- 69131 Ecully Cedex
marc-eric.bobillier@ec-lyon.fr

**RESUME**
Alors que la prise en compte des personnes handicapées dans la conception des technologies représente un enjeu social et politique important (loi du 11 février 2005 sur "l'obligation d'accessibilité numérique"), l'objectif de cette communication est d'apprécier le niveau d'accessibilité effectif de deux sites d'E-services à partir de tests d'usage et de préconiser des recommandations visant à améliorer l'accès pour le plus grand nombre.

**MOTS CLES :** Conception inclusive, accessibilité, utilisabilité, e-administration, personnes non-voyantes

**ABSTRACT**
While taking into account handicapped people in the design of technologies represents a social and political stake that becomes important (in particular with the recent law on equal rights for all the citizens, March 2004), this paper aims at evaluating the level of accessibility of two sites of E-services thanks to tests of use and proposing a set of recommendations in order to increase usability for the largest amount of people.

**CATEGORIES AND SUBJECT DESCRIPTORS:** H5.2 [User interfaces] Evaluation/methodology.

## INTRODUCTION ET CONTEXTE DE L'ETUDE

La tendance actuelle est de privilégier l'accès à l'information pour tous par voie électronique, en particulier pour les services administratifs et gouvernementaux (loi sur l'accessibilité numérique pour les E-services).

De nombreux projets de E-administration existent et connaissent un succès important en France (déclaration des impôts en ligne, consultation de données administratives : http://adae.pm.gouv.fr). Mais aujourd'hui, leur accessibilité n'est pas encore assurée [6]. En effet, malgré les efforts déployés par les gouvernements pour proposer des initiatives en ligne innovantes qui répondent aux besoins des citoyens, certains groupes de personnes demeurent à l'écart de ces transformations : ce sont toutes les personnes (âgées, handicapées, illettrées…) dont les limites et déficiences (cognitives, motrices, perceptives…) ne sont pas prises en compte lors de la conception de ces E-services. Or, l'accessibilité pour tous de ces sites devient un véritable enjeu social, culturel et politique pour l'intégration et la reconnaissance de ces publics à besoins spécifiques. La recherche ADELA que nous avons menée durant 12 mois avait pour objectif de faire un premier diagnostic sur les besoins et les niveaux d'accessibilité des E-services (http://www.adela.fr). Dans cette communication, nous exposerons les résultats qui portent principalement sur les difficultés d'usage que rencontrent les personnes aveugles lorsqu'ils interagissent avec des E-services, ainsi qu'un ensemble de recommandations pour pallier ces difficultés.

## L'ACCESSIBILITE DES TECHNOLOGIES AUX PERSONNES DEFICIENTES VISUELLES

Le consortium W3C (http://w3qc.org/docs/accessibilite.html) définit l'accessibilité par le fait que des personnes handicapées puissent "*percevoir, comprendre, naviguer et interagir de manière efficace avec le Web, mais aussi créer du contenu (texte, image, formulaire, son, …) et apporter leur contribution au Web*". Dans le champ des usages des nouvelles technologies, l'expression « personne handicapée » fait référence à une personne qui présente un usage particulier du dispositif, notamment sur la manière d'accéder et/ou de consulter l'information par rapport à la majorité des personnes.

On estime le nombre de déficients visuels à 1,2 million de personnes, dont 10% souffriraient de cécité totale [1]. Selon la gravité de cette déficience (malvoyant à aveugle), trois niveaux d'incapacité et des besoins en terme d'accessibilité peuvent être déclinés [1] et [4] : gêne légère dans l'accès à l'information (catégorie 1) ; difficulté de lecture du contenu (dégradation de la vue : catégorie

2) ; recours à une aide technique (logicielle et matérielle) pour accéder au contenu des pages (catégorie 3), tels les: lecteurs d'écran, les systèmes de synthèse/reconnaissance vocale ou encore les plages braille [10]. Du coup, un site deviendrait accessible dès lors qu'il serait assez flexible pour fonctionner avec [ces] différents dispositifs d'assistance et qu'il fournit un contenu accessible [9]

A mesure que la qualité des interfaces graphiques du Web s'améliore pour devenir plus utilisables et conviviales (*par des icônes, des onglets, des listes à cocher, des menus déroulants...*), on se rend compte que ces éléments graphiques représentent paradoxalement des obstacles à l'accessibilité pour les personnes aveugles [4] Afin d'être en mesure de prévenir et de corriger ce type de barrières, le W3C a développé, via son programme Web Accessibility Initiative (WAI), des normes d'accessibilité pour les sites Web. Il s'agit de 14 recommandations regroupant 60 sous-critères qui permettent d'évaluer la conformité d'un site aux directives d'accessibilité (http://www.w3.org/TR/2005/WD-WCAG20-20050630/checklist.html) Ces directives ont été reprises par des projets de loi sur l'accessibilité des sites web administratifs (section 508 aux Etats-Unis par exemple) et ont également inspiré des méthodes d'évaluation et de labellisation de l'accessibilité des sites : Blindsurfeur en Belgique, See it Right en GB [8] En France, Braillenet propose le label Accessiweb (http://www.braillenet.org/accessibilite) qui comptent 14 recommandations générales et 92 points à contrôler pour définir trois niveaux de qualité d'un site (bronze, argent et or), selon le niveau de respect des recommandations (priorité 1, 2 et/ou 3 comme pour le WAI). Mais la mise en œuvre de cette check-list de directives se révèle souvent longue et fastidieuse, donc lente et chère. Pour ces raisons, des outils d'inspection automatisée ont été élaborés à partir des critères du WAI afin d'identifier les problèmes d'accessibilité : Bobby, Ramp, Infocus, A-Promptt [7]. Les différentes enquêtes [1], [3], [5] et [6] utilisant ces outils indiquent d'ailleurs le très faible pourcentage sites WEB réellement accessibles : de 30 % à 2 % selon la catégorie du site (publics, gouvernementaux, commerciaux.

Remarquons enfin que la plupart des solutions d'accessibilité contribuent à "la conception universelle" (ou "conception pour tous") [12] qui bénéficient autant aux utilisateurs non-handicapés qu'à ceux qui présentent des incapacités.

## PRESENTATION DE LA METHODE UTILISEE

L'objectif est de comparer les qualités ergonomiques de deux sites de E-services (site de l'ANPE et de la Mairie de Vandoeuvre-les-Nancy) du point de vue des personnes valides et de celui des personnes aveugles. Il s'agit alors : (i) D'avoir une sorte de référentiel en terme d'usage afin d'appréhender les différentes stratégies de navigation sur le site et d'interpréter les problèmes rencontrés, (ii) De savoir si ces problèmes sont plus liés aux personnes, aux handicaps ou aux sites. Dans cette perspective, notre démarche consiste à effectuer (i) une inspection ergonomique des sites sur la base des critères de l'ergonomie des logiciels et de ceux de l'accessibilité (accessiweb) et (ii) des tests utilisateurs sur la base de trois scénarii de type différents (informationnel, interactionnel et transactionnel), et sur deux types de population (valides et aveugles).

### *Choix de l'Echantillon.*
8 personnes valides et 10 personnes aveugles ont participé à l'étude**.** Ces deux groupes présentaient des caractéristiques socio-biographiques équivalentes (âge, formation, sexe…). Pour les personnes valides, seule la maîtrise de l'Internet les distinguait alors que chez les personnes aveugles s'ajoutait le niveau de maîtrise du logiciel de lecture d'écran. Chacun des deux groupes était constitué de 5 experts et de 5 novices. Rappelons que selon Nielsen [11], des tests menés avec 5 utilisateurs suffisent à lever au moins 80 % des problèmes d'utilisabilité.

### *Méthode de Recueil des Données*
Les tests d'utilisabilité ont été effectués sur la base de trois scénarii couvrant le panel des E-services proposés : recherche d'information (Scénario 1), participation à un forum citoyens (Scénario 2) et remplissage d'une formulaire en ligne (Scénario 3). Afin d'obtenir un maximum d'informations, différentes techniques de recueil ont été utilisées :
• Verbalisation simultanée : "*penser à haute voix*" durant la réalisation du scénario.
• Des observations directes (par grille d'observations) et "indirectes" (par logiciel de capture d'écran (ViewLetCam) et caméra).
• Des questionnaires post-test de satisfaction (repris et adaptés de la grille Wammi, http://www.wammi.com/using.html)

A partir de ces différents résultats, nous avons cherché à interpréter les écarts de "performances" entre les personnes valides et aveugles afin de comprendre l'origine des difficultés éprouvées par ces dernières.

## DIFFICULTES D'USAGE ET RECOMMANDATIONS
Trois types de difficultés d'usage ont été identifiés :

### Difficultés liées à la Maîtrise d'Internet et à l'Utilisation du Logiciel d'Assistance Jaws

*Répétition des Barres de Navigation* : la personne aveugle parcourt systématiquement les mêmes barres de navigation pour chaque nouvelle page visitée (ce qui accroît sa charge mentale) alors que l'usager valide peut en faire abstraction en dirigeant son regard directement vers le contenu de la page. Or, Jaws propose une fonctionnalité -non exploitée par les usagers novices- qui permet d'éviter les barres de navigation pour atteindre le cœur de la page.

***Problèmes de Guidage*** : Pour assurer un guidage correct de l'utilisateur, il est important de lui indiquer les pages qu'il a déjà visitées. Pour les personnes aveugles, il suffirait d'activer une option de Jaws pour distinguer un lien déjà visité, option non connue par certains utilisateurs.

***Représentation de l'Organisation du Site :*** Le temps d'exploration des personnes aveugles est nettement plus grand que celui des valides pour les 3 scénarii, avec un écart type très grand chez les aveugles quelque soit le scénario envisagé. Cela peut s'expliquer par les différences de niveau dans la maîtrise des sites puisque les experts aveugles ont un temps de réalisation du scénario plus faible que celui des novices (43,3 secondes pour les experts contre 84,3 pour les novices). Il semble donc que ces experts détiennent des schèmes d'interaction qu'ils répliquent, avec plus ou moins de réussite, entre chaque site comme on a pu le constater dans le scénario 3 pour le remplissage du formulaire. "*Là, je crois bien que c'est l'adresse qu'on va me demander normalement ici*".

|   | Scénario 1 | | Scénario 2 | | Scénario 3 | |
|---|---|---|---|---|---|---|
| ***Moyenne*** | Val *105* | Av 814 | Val *229* | Av 1133 | Val *334* | Av 1176 |
| ***Ecart-Type*** | *85* | 598 | *145* | 381 | *277* | 1150 |

**Tableau 2** : Moyennes des temps d'exploration (exprimé en secondes), par catégorie d'usagers et scénario.

Il est donc nécessaire de proposer une formation adaptée à l'usage du Web et des dispositifs d'assistance (par exemple, "Jaws") pour les personnes aveugles. On a pu en effet constater que les difficultés d'utilisation provenaient de méconnaissance dans le paramétrage de ces systèmes d'aide.

**Problèmes Généraux de Conception des Sites Web**

***Densité Importante d'informations sur une Page :*** Les observations indiquent que tous les utilisateurs (valides et aveugles) sont gênés par la grande densité d'informations présentes sur les pages (par exemple 84 liens sur la page d'accueil de la mairie de Vandoeuvre-les-Nancy). Afin de ne pas submerger d'informations l'utilisateur (surtout pour les aveugles), il convient de ne pas surcharger les pages, en particulier la page d'accueil, en limitant le nombre de liens et en évitant les images décoratives inutiles qui « polluent » la lecture et nuisent à la lisibilité du site. Les informations doivent être organisées à l'écran selon leur importance et leur portée. Les informations pertinentes et cruciales doivent être positionnées en début de page pour être lues rapidement.

***Problèmes de Navigation :*** il est nécessaire de conserver la cohérence des barres de navigation tant dans leur structure que dans leur contenu et d'avertir l'utilisateur, par une alerte sonore, par exemple, lors de l'apparition dynamique des sous-menus. De la même façon, il est important de désactiver le lien de la page en cours alors que l'utilisateur la consulte. Les personnes aveugles (plutôt novices) ont souvent l'impression que leur choix n'a pas été pris en compte ou qu'elles se sont trompées. Elles n'ont donc pas le sentiment de contrôler le site et ont l'impression de tourner en rond.

***Lecture des Images :*** Les principales difficultés rencontrées par les personnes aveugles portent sur les images non commentées (sans texte alternatif) ou des images décomposées en sous-images qui donnent également lieu à une redondance de liens lors de l'exploration de la page par les dispositifs d'assistance.

***Liens Javascript*** : la présence de javascript (liens de navigation ou bouton de validation) bloque aussi l'utilisation des sites par les assistances techniques**.**

De mauvais choix techniques rendent ces sites administratifs inaccessibles. Il s'avèrerait donc opportun d'étudier les moyens de faire connaître et de faire appliquer, par les développeurs qui travaillent dans les entreprises ou qui créent des produits grand public, les normes et les principes qui permettent l'accessibilité des personnes déficientes visuelles aux applications. C'est un préalable nécessaire pour élaborer des dispositifs technologiques utilisables pour le plus grand nombre.

**Difficultés Spécifiques aux Sites Administratifs**

***Compréhension des Libellés :*** les personnes éprouvent des difficultés de compréhension face à des libellés d'items inappropriés, redondants ou polysémiques (par exemple, confusion entre « Téléservices » et « Téléprocédures » pour des internautes peu aguerris aux démarches administratives).

***Lecture des Tableaux*** : l'utilisateur prend connaissance une seule fois, en début de lecture, de l'intitulé des colonnes. Il doit alors faire l'effort de mémoriser le nom des colonnes pour parcourir le tableau. Ces intitulés devraient être énoncés avant la lecture du champ.
De plus. il est conseillé de réduire au maximum la verticalité des pages sous forme de tableau en organisant différemment les informations. A titre d'exemple, la page d'accueil du site de la mairie de Vandoeuvre-les-Nancy est construite sous forme de tableau. Comem le montre la figure 1, Jaws passe alternativement entre plusieurs rubriques d'informations (par des "va et vient") ; ce qui oblige l'utilisateur à reconstruire cognitivement la cohérence de l'ensemble.

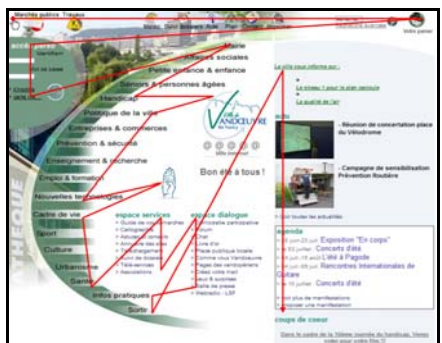

*Figure 1* : Page d'accueil du site de la mairie de Vandoeuvre-les-Nancy

***Structuration et Saisie des Formulaires*** : les utilisateurs aveugles se sont heurtés (dans le scénario 3) à la saisie du formulaire avec des difficultés pour passer du mode lecture au mode saisie. Pour faciliter la saisie de formulaires, il faut veiller à lier chaque étiquette de saisie à une étiquette « label » pour faciliter ce passage. Il faut également éviter de mettre des libellés, (informations, renseignements, explications, exemples) après les champs de saisie. De plus, il faut s'efforcer de rendre les formulaires les plus concis et les plus clairs possibles, pour éviter des peurs, ou des doutes sur la validité des informations saisies , tels en témoigne les commentaires d'un usager aveugle : "J*e vais tout relire car je me mélange, il y a tellement de trucs*".

***Difficultés d'Accès Direct aux Fonctions Clefs duS-site*** : compte tenu des sources d'informations très nombreuses et variées proposées à l'usager aveugle, nous préconisons la possibilité d'accéder directement à aux rubriques clefs du site (recherche, plan du site…. mais aussi téléprocédures, formulaire de contact…) par des raccourcis claviers communs à tous les sites administratifs (par exemple, "Alt+ Ctrl T" pour accéder directement aux télé-procédures …).

## CONCLUSION

L'objectif de cet article était d'évaluer le niveau d'accessibilité des E-services. Il ressort que les problèmes d'usage viennent principalement d'un manque de maîtrise du Web et des dispositifs d'assistance par les usagers aveugles, d'un manque de prise en compte des principes d'accessibilité dans la conception des sites et de E-services insuffisamment adaptés aux personnes aveugles. Si donc la nécessité d'une e-administration accessible est clairement posée, le schéma global des moyens et solutions technologiques et de sensibilisation (tant aux usagers qu'aux concepteurs pour développer des e-services accessibles) n'est pas aujourd'hui clairement identifié. La demande sociale est certes soutenue par la volonté politique (loi sur l'accessibilité numérique principalement) mais la méthodologie de réalisation effective est encore à définir, notamment chez les développeurs qui semblent peu concernés par ces enjeux [9].